# Coulombic efficiency and capacity retention are not universal descriptors of cell aging

Marco-Tulio F. Rodrigues

Chemical Sciences and Engineering Division, Argonne National Laboratory, Lemont, IL, USA

**Contact:** marco@anl.gov

**Abstract**

Capacity and coulombic efficiency are often used to assess the performance of Li-ion batteries, under the assumption that these quantities can provide direct insights about the rate of electron consumption due to growth of the solid electrolyte interphase (SEI). Here, we show that electrode properties can actually change the amount of information about aging that can be directly retrieved from capacity measurements. During cycling of full-cells, only portions of the voltage profiles of the cathode and anode are accessible, leaving a reservoir of cyclable $Li^+$ stored at both electrodes. The size and availability of this reservoir depends on the shape of the voltage profiles, and accessing this extra $Li^+$ can offset some of the capacity that is consumed by the SEI. Consequently, capacity and efficiency measurements can, at times, severely underestimate the rate of side reactions experienced by the cell. We show, for example, that a same rate of SEI growth would cause faster capacity fade in $LiFePO_4$ than in NMC cells, and that the perceived effects of aging depend on testing variables such as depth of discharge. Simply measuring capacity may be insufficient to gauge the true extent of aging endured by Li-ion batteries.

**Introduction**

Li-ion batteries will constantly experience parasitic processes that lead to irreversible consumption of electrolyte constituents.[1-3] Much of the performance fade observed in cells can be attributed to the reduction of electrolyte species at the surface of the negative electrode (NE), forming an imperfect passivation layer known as solid electrolyte interphase (SEI). The slow but continuous progression of this "SEI growth" will deplete electrons out of the inventory supplied by the positive electrode (PE), generally leading to capacity loss. Furthermore, it will also lead to electrolyte consumption,[4] starving the cell off its charge-conducting medium. Much effort is dedicated to identifying cell formulations and operating conditions that minimize the extent of aging processes that may be experienced by a Li-ion battery. These efforts often involve analyzing capacity and coulombic efficiency (CE) data, which are expected to provide reliable information about electrolyte degradation levels along the cell life. The present work will discuss whether this expectation – that cell health is accurately measured by capacity and CE – will always hold true.

The *measurability* of aging is known to be affected by the shape of the voltage profiles of the positive and negative electrodes at the end of discharge (EOD) of the full-cell.[5, 6] Consider Figure 1, which shows the voltage profiles of a LiFePO$_4$ (LFP) PE and a graphite (Gr) NE during the discharge of two hypothetical full-cells. In these figures, the x-axis indicates the state-of-charge (SOC) of the full-cell, and thus portions above 1 and below zero are not actively utilized during the half-cycle being shown. The main distinction between Figure 1a and Figure 1b lies in the relative behavior of the PE and the NE at the EOD (i.e., at SOC = 0 in this example). In Figure 1a, the full-cell will meet the EOD when the voltage profile of the NE exhibits a strong upward trajectory. Qualitatively, this can be thought of as involving a near-complete Li$^+$ extraction from the anode while the cathode still retains empty sites when the cell reaches the EOD.[6] Conversely, in Figure 1b the LCV is met when the PE undergoes a rapid decrease in potential; here, all available sites of LFP are replenished with $Li^+$ but there is still substantial capacity stored at the anode. The first case (Figure 1a) illustrates a cell in which discharge is *limited by the NE*. Figure 1b represents the case with discharge being *limited by the PE*. Note that the *charge* half-cycle is PE-limited in both cases. Most conventional Li-ion batteries present NE-limited discharge, with the empty sites in the PE being a consequence of the capacity lost due to cell aging and initial SEI growth. PE-limited cells can exist due to "prelithiation" of the negative electrode (as commonly done in Si-containing cells, refs. [7-9]) or due to PE degradation while the cell is in the charged state.[10-12] This "prelithiated" quality of PE-limited cells is useful to illustrate how it can affect the measurability of aging. When SEI growth happens in a PE-limited cell, the excess $Li^+$ *reservoir* that is left in the NE (i.e., when cell SOC < 0 in Figure 1b) can compensate for the $Li^+$ that is lost to reduction side reactions, causing little or no capacity loss to actually be measured.[6] That is, the parasitic processes that are normally responsible for capacity fade are still happening, but their effect is not directly measurable. On the other hand, at an ideal NE-limited cell, the net electron loss to side reactions can be directly measured as capacity fade.[6] While NE-limited cells (Figure 1a) allow a nearly perfect measurability of the consequences of side reactions, the information provided by PE-limited cells (Figure 1b) is close to none.

Although the examples shown in Figure 1 are relevant to discussing the interpretation of cell behavior, they only represent the extremes of a spectrum of states in which cells can exist. *Exploring how measurement outcomes depend on these states is the focus of the present work.* Consider Figure 2a, which exhibits voltage profiles during discharge of a hypothetical cell containing a $LiNi_{0.8}Mn_{0.1}Co_{0.1}O_2$ (NMC811) PE and a NE containing 70% $SiO_x$. Inspection of the slopes of the voltage profiles of the positive and negative electrodes at the EOD (SOC = 0) shows that they are significantly more similar than in the cases of Figure 1. In other words, the discharging of this cell is not strongly limited by either electrode.[5] Another interesting scenario is shown in Figure 2b for a hypothetical NMC811 vs. Gr cell. Although the cell is NE-limited when fully discharged (SOC = 0), it becomes PE-limited when discharge is interrupted at most other points (as in SOC = 0.1, indicated by dashed lines), as the EOD would occur at a plateau of Gr. Incomplete discharge is common in many applications of Li-ion batteries (such as electric vehicles), and many studies incorporate "shallow" cycling when evaluating aging behavior.[13, 14]

Intuitively, one would expect that the case shown in Figure 2a, in which discharge is not strongly limited by either electrode, would display a behavior that is intermediate to the discussed for Figure 1a and Figure 1b. In that case, *some* of the SEI growth can be directly inferred from capacity measurements but not all of it. Next, we review how this information gap can be quantified.

***Measurability of aging.*** If we disregard impedance effects and the loss of accessible active material capacity, cell aging is essentially the consequence of reduction and oxidation side reactions taking place at the negative and positive electrodes, respectively. Reduction side reactions generally involve the consumption of electrons at the NE to form SEI. Oxidation side reactions involve the transference of electrons from the electrolyte and/or surface layers to the PE. Hence, while in the former case the parasitic process will *decrease* the initial cyclable electron inventory of the cell, the latter will cause an *increase*.[6] While oxidation side reactions can result in temporary capacity gain, it generally involves the decomposition of electrolyte species, in which case it will inevitably contribute to detrimental effects, such as electrolyte depletion, if allowed to proceed unchecked for extended periods.

Consider that the time-averaged rate of reduction and oxidation parasitic processes within a given cycle are expressed as $I_{red}$ and $I_{ox}$, respectively. We will use these quantities that indicate parasitic currents to express the cycle-to-cycle capacity retention (CR, measured with respect to the preceding discharge half cycle) and coulombic efficiency (CE). Tornheim and O'Hanlon have previously demonstrated (ref. [6]) that, for the ideal case in which cell charging is strongly limited by the PE and discharging is limited by the NE (Figure 1a), CE and CR can be expressed as

$$CE = \frac{I - I_{red}}{I + I_{red}} \qquad \text{(eq. 1)}$$

$$CR = \left(\frac{I - I_{red}}{I + I_{red}}\right)\left(\frac{I + I_{ox}}{I - I_{ox}}\right) \qquad \text{(eq. 2)}$$

Here, $I$ indicates the constant current at which the cell is cycled, and it is assumed that $I$, $I_{red}$ and $I_{ox}$ remain constant during charge and discharge. The equations above indicate that, for NE-limited cells, coulombic efficiency values solely inform about the extent of SEI growth and are thus unaffected by oxidation side reactions. Meanwhile, CR depends on the balance between capacity loss (from $I_{red}$) and capacity gain (from $I_{ox}$).

For the ideal case in which both charge and discharge are limited by the PE (Figure 1b), Tornheim and O'Hanlon have shown (ref. [6]) that CE and CR are given by

$$CE = \frac{I - I_{ox}}{I + I_{ox}} \qquad \text{(eq. 3)}$$

$$CR = 1 \qquad \text{(eq. 4)}$$

As discussed above for the case of prelithiation, this type of cell does not exhibit measurable capacity fade, even for finite values of $I_{red}$; cell capacity would remain steady until the Li$^+$ reservoir is depleted and the cell is no longer PE-limited. Quite interestingly, CE measurements in a PE-limited cell will inform only on the extent of oxidative parasitic processes, and thus cannot convey any information about SEI growth. This unexpected point becomes clear once one considers that the transference of electrons to the PE (from $I_{ox}$) will affect when the cell reaches the ending of both charge and discharge half-cycles, as they are dictated by strong polarization of the PE. The same does not apply to SEI growth, since it occurs at the NE and has no direct effect on the Li$^+$ content of the PE. The cell essentially behaves as if it were a half-cell (i.e., PE vs. Li metal).

In a previous work,[5] we have shown that the equations above can be generalized by explicitly considering how side reactions will affect the potentials of PE and NE at the end of each half-cycle. It was demonstrated that

$$CE = \frac{[I - I_{red}(1-\lambda) - \lambda I_{ox}]}{[I + I_{red}(1-\lambda) + \lambda I_{ox}]} \quad \text{(eq. 5)}$$

$$CR = CE \frac{[I + I_{ox}(1+\omega) - \omega I_{red}]}{[I - I_{ox}(1+\omega) + \omega I_{red}]} \quad \text{(eq. 6)}$$

Here, CE and CR depend on both $I_{red}$ and $I_{ox}$, to an extent given by the weights $\lambda$ and $\omega$. These two parameters carry information about the degree at which the PE and NE are limiting charge and discharge, with

$$\lambda \equiv \frac{\left.\dfrac{dU_{pos,d}}{dq}\right|_{LCV}}{\left(\left.\dfrac{dU_{pos,d}}{dq}\right|_{LCV} - \left.\dfrac{dU_{neg,d}}{dq}\right|_{LCV}\right)} \quad \text{(eq. 7)}$$

$$\omega \equiv \frac{\left.\dfrac{dU_{neg,c}}{dq}\right|_{UCV}}{\left(\left.\dfrac{dU_{pos,c}}{dq}\right|_{UCV} - \left.\dfrac{dU_{neg,c}}{dq}\right|_{UCV}\right)} \quad \text{(eq. 8)}$$

In the equations above, UCV and LCV denote upper and lower cutoff voltages, respectively, indicating the condition in which the cell meets the end of cell charge (c) and discharge (d). The derivatives are the "terminal" slopes of the voltage profiles of PE (pos) and NE (neg) at the end of the appropriate half-cycle. Note that the sign of the slopes depends on the type of half-cycle being considered; e.g., the slope of the NE is positive at the LCV (see Figure 1a, for example) but negative at the UCV. With that in mind, it can be observed that $\lambda$ is a positive number but $\omega$ is negative.

The coefficient $\lambda$ describes the "limitation" of the discharge half-cycle, while $\omega$ describes the same for charge. Returning to the examples in Figure 1, $\lambda = 1$ for a PE-limited discharge and $\lambda = 0$ when it is limited by the NE (see Table 1). If the cell charging is limited by polarization of the PE (i.e., if the magnitude of the terminal slope of the PE is much larger than that of the NE) and discharging is limited by the NE, equations 5 and 6 revert of the limiting cases derived by Tornheim and O'Hanlon in equations 1 and 2. A similar exercise can be done for the PE-limited case. Crucially, equations 5-8 also allow us to describe cases such as the illustrated in Figure 2a, by computing the slopes of each electrode when neither of them is strongly limiting cell discharge.

Furthermore, it can also describe cases of interest in which cell charging may not be limited by the PE.

Alternatively, we have also shown (ref. [5]) that equation 6 can be re-written as

$$CR \approx \frac{I - (I_{red} - I_{ox})(1 + \omega - \lambda)}{I + (I_{red} - I_{ox})(1 + \omega - \lambda)} \qquad \text{(eq. 9)}$$

where capacity retention is expressed as a function of the net parasitic current $(I_{red} - I_{ox})$. This formulation has two advantages. One is that the net parasitic current can be directly correlated with the slippage $(q_{red} - q_{ox})$ caused by aging, where $q_i$ indicates the charge exchanged by the appropriate side reaction within the period of interest. Slippage (also known as changes in offset), is the alteration of the relative "alignment" of the voltage profiles of PE and NE as a result of parasitic processes.[5, 9, 12] This quantity can be determined experimentally and used to gauge the occurrence of side reactions even when they do not produce measurable capacity fade,[5] and is thus a robust descriptor of aging. The second advantage of using equation 9 is that it conveniently codifies the measurability of aging in the *information factor* $(1 + \omega - \lambda)$. When $1 + \omega - \lambda = 1$, the consequences of side reactions can be accurately measured as cell capacity fade (like in Figure 1a). When $1 + \omega - \lambda = 0$, no information is available, like discussed for Figure 1b. Other possible scenarios, including the case shown in Figure 2a, can also be quantified with this equation.

In summary, the contributions of $I_{red}$ and $I_{ox}$ to coulombic efficiency are dictated by the value of $\lambda$, which relates to the terminal slopes of voltage profiles of the positive and negative electrodes at the EOD. The type of information conveyed by CE measurements will depend on the instantaneous value of $\lambda$, and can vary widely depending on the active materials and testing procedure. For capacity retention, whether or not side reactions will produce measurable outcomes will depend on the quantity $(1 + \omega - \lambda)$, which considers the terminal slopes at both end of charge (EOC) and EOD.

In the present work, we apply this formalism to discuss situations that are commonly encountered in battery science and that involve marked changes in the measurability of aging. We show that, all things constant, the exact same rate of aging will produce substantially lower CE and CR values when LFP is used as PE instead of NMCs. Additionally, the information provided by capacity measurements is shown to vary widely with the specified cutoff voltages of charge and discharge, and with cycling rate. Abstracting information about the true extent of aging of a cell out of testing data may require additional layers of reasoning. The present work attempts to raise awareness for this fact, and to provide tools that help quantifying this aging behavior.

*Experimental*

The hypothetical cells discussed here were simulated by combining voltage profiles for the specified PE and NE at arbitrary negative/positive (N/P) ratios. The voltage profiles were obtained experimentally from half-cells (vs. Li metal) cycled at rates < C/100; this data has been observed to reproduce well the behavior of electrodes in full-cells that are cycled at slow rates (C/10 – C/25).[10] While some of the quantitative considerations in this work may depend on specific parameters of the simulated full-cells (such as N/P ratio and PE/NE offset), the examples provided here are realistic and relevant to battery science, providing useful illustration of expected trends.

Half-cell data was obtained in 2032-format coin cells, using an electrolyte consisting of 1.2 M LiPF$_6$ in ethylene carbonate and ethylmethyl carbonate (3:7 wt:wt), procured from Tomiyama. When testing silicon-based electrodes, the electrolyte further contained 3 wt% of fluoroethylene carbonate (Solvay). All electrodes were fabricated at Argonne National Laboratory's Cell Analysis, Modeling and Prototyping (CAMP) Facility, and were composed of the active material, C45 carbon additive (Timcal) and a binder. Specific information for each electrode is provided below. The indicated porosity was achieved after calendering the electrode to the indicated coating thickness.

- NMC811: 90% active (Targray), 5% C45, 5% PVDF (5130, Solvay); 34.5% porosity, 59 µm, 15.81 mg/cm$^2$
- NMC532: 90% active (Toda), 5% C45, 5% PVDF (5130, Solvay); 33.1% porosity, 42 µm, 11.4 mg/cm$^2$
- LFP: 90% active (Johnson Matthey), 5% C45, 5% PVDF (5130, Solvay); 38.6% porosity, 95 µm, 19.16 mg/cm$^2$
- Graphite: 91.83% active (SLC1506T, Superior Graphite), 2% C45, 0.17% oxalic acid, 6% PVDF (KF-9300, Kureha); 37.4% porosity, 47 µm, 6.49 mg/cm$^2$
- SiO$_x$: 70% active (Osaka Titanium Technologies Co.), 10% C45, 20% polyimide binder (P84, HP Polymer GmbH); 45% porosity, 20 µm, 2.18 mg/cm$^2$

Graphical determination of simulated measurable capacity fade as a function of electrode slippage was performed using the 'Alawa toolbox.[12, 15, 16] 'Alawa simulates voltage profiles of full-cells using half-cell data as input, enabling the analysis of the expected effects of various degradation modes. SEI growth was simulated by imposing a constant rate of lithium inventory loss (LLI) to NMC vs. graphite cells operating within different voltage cutoffs. The UCV was always 4.1 V, while the LCV varied to match depths of discharge of 1, 0.8, 0.6 and 0.4 in the unaged cell. In the formalism of the present work, LLI is equivalent to the net parasitic charge ($q_{red} - q_{ox}$). These simulations used voltage profiles provided as part as 'Alawa's library.

*Results and Discussion*

***How the positive electrode affects the measurability of aging.*** In a previous work, we have discussed in detail how the shape of the voltage profile of the NE affects the outcomes of CE and CR measurements, highlighting the fact the capacity and CE measurements largely underestimate the rate of SEI growth in silicon-containing electrodes.[5] Here, we will extend a similar analysis to PE active materials.

Consider the hypothetical cells shown in Figure 3. The cells can contain one of three types of PE (NMC811, NMC532 and LFP), and either Gr or $SiO_x$ as NE. Voltage profiles exhibited during cell charge and discharge are displayed in Figure 3a and Figure 3b, respectively. Corresponding differential voltage (dV/dQ) profiles are shown in Figures 4a-b. Taking these values at EOC/EOD for all possible electrode pairs and applying them in equations 7 and 8, the parameters $\lambda$ and $\omega$ can be computed (Figures 4c-f). Analyzing Figure 3, Figure 4 and equations 7 and 8, the following heuristics can be found: i) if the PE presents a large terminal slope during charge (such as LFP), $\omega \to 0$ and charge is strongly PE-limited; ii) when the NE has a large slope at the EOD (such as Gr), $\lambda \to 0$ and discharge is strongly NE-limited; iii) NMC cells paired with a Si-based NE will present larger magnitudes for $\lambda$ and $\omega$ (that is, electrode limitation at either half-cycle is weaker than in Gr cells); iv) the "flatness" of LFP at the EOD and its sharp potential rise at the EOC decreases the role of the NE in affecting measurement outcomes, if compared with NMC cathodes.

Determination of the values for parameters $\lambda$ and $\omega$ in the various cells allows us to use equations 5 and 9 to infer how the identity of the electrode will affect the information that is conveyed by CE and CR measurements. For this analysis, we assume that all cells experience the same rates of side reactions (i.e., a same $I_{red}$ and $I_{ox}$); for CE, we further assume that $I_{ox} = 0.5$ $I_{red}$, as estimated experimentally elsewhere for a NMC vs. Gr cell.[6] The resulting projected values for CE and CR are shown in Figure 5; CR values are plotted as a function of the net parasitic current ($I_{red} - I_{ox}$) normalized by the cycling current $I$ (see equation 9). When Gr is the NE, the measurable coulombic efficiency values are nearly independent of the PE (Figure 5a). Since $\lambda$ is close to zero in all cases, these cells approach the ideal NE-limited scenario described by Tornheim and O'Hanlon in equation 1, and coulombic efficiency conveys reliable information about SEI growth.[6] However, differences between the PEs at the EOC cause sufficient changes to $\omega$ (Figure 4c) to affect capacity retention (Figure 5b): cells using LFP would fade somewhat faster than their NMC counterparts simply because the voltage profile of the PE has a different shape. To put the numbers of Figure 5b in context, if we assume that all cells would display a measurable CR of 0.9995 (dashed line), the net parasitic current ($I_{red} - I_{ox}$) in NMC cells would be a remarkable *14% higher* than when the cathode is LFP.

If the NE is $SiO_x$, the coulombic efficiency measured in LFP cells will be lower than with NMC for a same underlying rate of aging (Figure 5c). Using the values of $\lambda$ shown in Figure 4f, equation 5 can be used to express CE for all cathodes as

$$CE_{LFP} = \frac{[I - 0.98 I_{red} - 0.02 I_{ox}]}{[I + 0.98 I_{red} + 0.02 I_{ox}]} \qquad (\text{eq. 10})$$

$$CE_{532} = \frac{[I - 0.67 I_{red} - 0.33 I_{ox}]}{[I + 0.67 I_{red} + 0.33 I_{ox}]} \qquad (\text{eq. 11})$$

$$CE_{811} = \frac{[I - 0.60 I_{red} - 0.40 I_{ox}]}{[I + 0.60 I_{red} + 0.40 I_{ox}]} \qquad (\text{eq. 12})$$

Coulombic efficiency measurements are good descriptors of SEI growth in cells with a high SiO$_x$ content only when LFP is the PE. We have previously suggested the use of this positive electrode for experiments that intend to probe SEI stability of Si-based NEs.[5] For the NMC positive electrodes, CE measurements will also convey information about the extent of oxidation side reactions, especially for NMC811. Since cells generally tend to *lose* capacity over time and testing, $I_{red} > I_{ox}$, and the larger weight of the smaller number ($I_{ox}$) in equations 11 and 12 leads to an increase in CE values relative to LFP cells. The gap between NMCs and LFP is even larger when we analyze capacity retention (Figure 5d). Using the parameters shown in Figures 4e-f, equation 9 becomes

$$CR_{LFP} \approx \frac{I - 0.98(I_{red} - I_{ox})}{I + 0.98(I_{red} - I_{ox})} \qquad \text{(eq. 13)}$$

$$CR_{532} \approx \frac{I - 0.54(I_{red} - I_{ox})}{I + 0.54(I_{red} - I_{ox})} \qquad \text{(eq. 14)}$$

$$CR_{811} \approx \frac{I - 0.47(I_{red} - I_{ox})}{I + 0.47(I_{red} - I_{ox})} \qquad \text{(eq. 15)}$$

From the equations above, it can be seen that CR measurements in LFP cells convey significantly more information about aging than when an NMC-based PE is used. Consequently, LFP and NMC cells experiencing the same rate of parasitic processes will lead to a steeper capacity fade with LFP systems than in the latter case. In this case, measuring a CR of 0.9995 (dashed line in Figure 5d) in a cell with NMC811 would conceal a net parasitic current $(I_{red} - I_{ox})$ that is *108% higher* than in a LFP cell. Differences observed among the types of NMCs are still significant, though less dramatic than noted with LFP.

A key consideration needed to compute the values exhibited in Figures 5a,c was that $I_{ox}$ was the same in all cases. Assuming that the rate of oxidation side reactions is proportional to the average PE potential, that would hardly be the case, as LFP operates at much lower potentials than NMCs (see Figure 3). Three points are worth highlighting here. The first is that equation 10 shows that CE depends weakly on $I_{ox}$ for LFP cells, making the assumption from Figures 5a,c relevant only for calculations in NMC cells. The second point relates to the fact that the x-axis in Figures 5b,d represents the net parasitic currents ($I_{red} - I_{ox}$). If $I_{ox}$ is generally smaller for LFP, that implies that a given vertical line in Figures 5b,d represents a case in which $I_{red}$ is smaller for LFP than for the NMCs (as a same net parasitic current results from subtracting a smaller number from a smaller number). Consequently, a same rate of SEI growth ($I_{red}$) would produce faster capacity fade in LFP cells than can be directly glanced from Figures 5b,d. Finally, an interesting study by Aiken et al. have provided a direct comparison between Gr cells with NMC532 tested a low cutoff voltages (3.65 and 3.80 V) and with LFP.[17] Under these conditions, $I_{ox}$ can be expected to present similar values for both PEs, making quantitative trends from Figure 5b more transferrable to this case. Their work reported that cells with NMC generally presented cycle lives that were superior to the ones using LFP, in agreement with the discussion above. While there may be additional factors contributing to the different cell performances observed with these cathodes, there are fundamental reasons to expect LFP to display lower CE and CR when the overall rates of aging are similar.

In summary, this section discussed how, all else constant, changing the PE used in a cell can lead to severe differences in the way the extent of parasitic processes can be directly gauged by capacity and CE measurements. This variance originates from differences in the shape of the voltage profiles of PE and NE at the end of charge and discharge half-cycles, and can also vary in relative magnitude depending on the choice of NE. Naturally, altering the active material in a Li-ion battery can have other effects that are not included in the analysis above, such as crosstalk, gassing, particle cracking and impedance rise.[1, 10, 11, 18] Nevertheless, this section also shows that *some* of the differences in perceived cell performance at varying PEs are intrinsic to the thermodynamic open circuit potential curve of each active material, which must be considered when analyzing aging behavior.

*How the depth of discharge affects the measurability of aging.* So far, most of the cases we analyzed involved cells that were fully discharged (i.e., until they return to SOC = 0). This case is defined as involving a depth of discharge (DOD) of 1. Figure 2b shows a case in which the EOD occurs when the cell is at SOC = 0.1 (that is, a DOD of 0.9), indicating how a cell with nominally NE-limited discharge (at a DOD of 1) could be become mostly PE-limited when operating at another condition. The present section focuses on analyzing this phenomenon.

Consider again all the possible cells represented in Figure 3. To analyze how variation in the DOD would affect the information that can be probed by CE and CR measurements, we assume that all cells were initially fully charged (to SOC = 1), and then discharged to DODs that varied from 0.3 to 1. The scenario where DOD = 1 was discussed in detail in Figure 4 and Figure 5. The values of $\lambda$ at different DODs can be obtained by considering the slopes of the PE and NE at different points of Figure 4b. Once these values are calculated for all PE/NE combinations, we can study the impact of DOD on coulombic efficiency (through $\lambda$, as in equation 5) and on capacity retention (through the information factor $[1 + \omega - \lambda]$, as in equation 9). These two quantities are shown in Figure 6 for the hypothetical cells of interest. Note that, since charging conditions remained constant, $\omega$ is invariant for each cell.

When $\lambda$ is close to zero, cell discharge is strongly NE-limited and CE can reliably convey the extent of SEI growth. Figure 6a shows $\lambda$ as a function of DOD for Gr cells containing the three types of PEs. In all cases, a full discharge leads to a strongly NE-limited case. However, most other conditions will cause EOD to be rather weakly limited by the NE. In fact, NMC cells can even become *PE-limited* in many instances (such as when DOD is 0.45-0.6, as $\lambda \rightarrow 1$). These changes in $\lambda$ carry important consequences. As discussed in the previous section, higher values of $\lambda$ imply on growing contributions of $I_{ox}$ to measurable CE. Hence, CE becomes a poorer source of information about SEI growth when a NMC vs. Gr cell is not fully discharged. Additionally, as discussed above, CE values are made higher in these conditions, since $I_{ox} < I_{red}$ (see equation 5). In this case, achieving higher coulombic efficiencies by constraining the DOD does not signify that the cell is more stable, just that the manifestation of side reactions on CE has changed. Gr cells with LFP will remain NE-limited unless the EOD occurs within a graphite plateau, in which case the voltage profiles of both electrodes are "flat" and the termination of discharge is no longer dominated by a single electrode.

Figure 6b exhibits the information factor $(1 + \omega - \lambda)$, which gauges how much the side reactions will cause measurable capacity fade (equation 9). Values closer to 1 imply a higher measurability. As we discussed above, an ideal NE-limited cell (i.e., with $\lambda \rightarrow 0$) has perfect measurability, and thus Figure 6b is the mirror image of Figure 6a (vertically shifted by $\omega$). Even at full discharge, CR in NMC vs. Gr cells cannot capture aging as well as a cell using LFP as the positive electrode. The measurability for the NMCs further deteriorates at lower DODs. Curiously, for these cells $(1 + \omega - \lambda)$ even becomes negative at certain DODs (< 0.35 and 0.45-0.65). In these cases, side reactions should actually cause the cell to show moderate *capacity gain* in the typical case when $I_{red} > I_{ox}$.

For cells using a SiO$_x$-rich negative electrode (Figures 6c-d), when LFP is the PE, CE will always be correlated with SEI growth and capacity measurements can track the net parasitic current quite well. For NMCs, CE and CR will always underestimate the rate of SEI growth. In this latter case, the effect of the DOD is much weaker than in graphite cells, due to the lack of sharp features in the voltage profile of the NE.

Equations 6 and 9 can only determine CR within a narrow range of cycles, as these expressions are only valid while the terminal slopes of the PE and the NE remain constant; i.e., $\lambda$ and $\omega$ are invariant. For a direct illustration of how DOD can affect capacity fade, we employed the `Alawa toolbox to calculate graphically the perceived effects of cell aging.[12] Figure 7a shows the discharge voltage profiles for a NMC PE (gold), graphite NE (purple) and the simulated full-cell profiles (black). Aging is simulated by imposing a relative lateral shift between the PE and NE profiles (slippage), as both oxidation and reduction side reactions will involve electron exchanges at a single electrode (PE and NE, respectively). Thus, when the PE has a given Li$^+$ content, the amount of capacity held by the NE will depend on the state of aging of the cell. In Figure 7a, we represent the shifts caused by losses of every 2% of the initial cyclable electron content of the PE, to a total of 10%. These shifts are equal to the net capacity consumed by parasitic processes, which is given by $(q_{red} - q_{ox})$. After determining the new "alignment" of the electrodes, profiles for the full-cell can be calculated (as $U_{cell} = U_{pos} - U_{neg}$). By constraining the simulated cell profile within the assumed UCV and LCV, the measurable capacity at a given DOD can be estimated. Note that slippage is a consequence of the net loss of electrons by the cell and that equal losses will cause equal slippage, regardless of the voltage window. As we discuss later, this property makes slippage a useful metric to quantify cell aging.

In Figure 7a, the constant UCV of 4.1 V is indicated by the gray dashed lines. The corresponding LCVs for DODs of 0.4, 0.6, 0.8 and 1 are also shown in the figure. Figure 7b shows the expected capacity fade vs. the actual capacity consumed by side reactions for the various DODs. The diagonal dashed line indicates the case in which the cell would display perfect measurability, in which case the net electron loss to side reactions could be accurately assessed through capacity measurements. Clearly, none of the simulated conditions can provide this level of accuracy, in agreement with the discussion about Figure 6b. Furthermore, the level of information about the state of aging of the cell that can be gathered from capacity measurements deteriorates rapidly as we decrease the DOD. At a DOD of 0.8, sudden changes in the trajectory can be seen in Figure 7b. The slope of these curves is proportional to $(1 + \omega - \lambda)$, and this nonlinearity indicates that terminal slopes have varied within the considered extent of slippage. Interestingly, a slight capacity gain (i.e., a negative fade) is observed for a DOD of 0.4, in agreement with the analysis of Figure 6b. Note that the cells discussed in Figure 6 and Figure 7 differed in their exact initial state (electrode profiles used in simulations, assumed N/P ratio etc.; see *Experimental* section), and this capacity gain may not occur at a same point for both examples.

This section shows that the concept of "electrode limitation" for a given half-cycle and its consequences to the measurability of aging is rather fluid. A same type of cell can exhibit behaviors that range from mostly NE- to PE- limited depending on the specific testing conditions. As we have shown, the magnitude of coulombic efficiency and its correspondence with the rate of SEI growth experienced by the cell can vary widely across these conditions. In some cases, cell capacity may even increase as a response to aging. Directly comparing the performance of cells that were tested under different conditions can be extremely misleading. We emphasize that this analysis does not consider mechanical effects associated with varying the depth of cycling, such as active material particle cracking that can lead to isolation.[19] Nevertheless, this discussion highlights that there are intrinsic reasons to expect that different testing conditions will lead to different measurable outcomes.

***The depth of charge and the cycling rate can also affect the measurability of aging.*** The reader should now be familiar with the fact that changes in the shape of the voltage profiles of the PE and the NE at the end of a half-cycle can affect the meaning of CE and CR. The examples above have also illustrated how such changes may occur as a consequence of altering the testing conditions to which a Li-ion battery is subjected. In this section we briefly extend this analysis to the effects of the depth of charge (DOC) and cycling rate. To avoid repetition, we will limit this discussion to highlighting the parameters of interest without including the complete discussion.

Figure 8a shows how the information factor $(1 + \omega - \lambda)$ will vary in NMC811 vs. Gr and LFP vs. SiO$_x$ cells as a function of the DOC. The depth of discharge is assumed to be 1 in all cases. The former cells present high (though different) values of the information factor for all DOCs but 0.55-0.6. This domain corresponds with the occurrence of a steep region in the voltage profile of Gr (Figure 3a), causing the cell charge to become more weakly limited by the PE. For the LFP vs. SiO$_x$ cell, all but a full charge will result in a scenario in which aging does not produce significant measurable capacity fade. Once again, comparing the performance of cells exposed to different testing conditions can be complex, as aging will cause different effects on measurable CE and CR.

The effects of rate can be quite complex. Consider Figure 8b, which exhibits NE profiles measured with a reference electrode as a full-cell (vs. NMC532) is discharged at different rates following a full charge. While at C/10 the NE presents a large slope at the EOD (implying a likely

NE-limited condition), the end portion of the profiles become much smoother at higher currents. Furthermore, the plateaus associated with the evolution of $LiC_x$ phases completely vanish at 2C and above, changing how variations in DOD would affect CE and CR values. Rate intrinsically will affect these quantities (as equations 5, 6 and 9 depend explicitly on the cycling current *I*) but these changes in the shape of the voltage profile cause the NE to behave as a different material altogether.

*Bypassing the reservoir effect.* In a previous work, we discussed that capacity and CE measurements will largely underestimate SEI growth in cells containing large quantities of silicon-based materials in the negative electrode.[5] Qualitatively, this phenomenon was explained by the *reservoir effect*. The voltage profile of many PE materials of commercial interest displays an increase in electrode potential at decreasing $Li^+$ content (that is, the PE potential is raised during cell charging). When capacity is lost to side reactions, the remainder $Li^+$ is insufficient to restore the PE to its previous EOD potential, causing the terminal potential experienced by the positive electrode when the cell is discharged to increase; this is commonly observed using a reference electrode.[9, 20] Assuming that the EOD occurs at a fixed full-cell voltage, this increase in PE potential will force an equal change in the terminal potential of the NE. The key realization of our previous work was to recognize the consequences of the fact that raising the terminal NE potential causes additional delithiation of the negative electrode.[5] All negative electrodes will hold some amount of $Li^+$ even when the nominal *cell SOC* is zero, and depletion of this *reservoir* as the terminal NE potential increases can distort the correlation between measurable capacity fade and the actual capacity that is consumed by side reactions. The rate of release of the reservoir is inversely proportional to the instantaneous slope of the NE voltage profile at the EOD (for a fixed cathode), causing this effect to be larger for Si than for graphite (compare NE profiles in Figure 2, for example).[5] Having capacity measurements underestimating SEI growth can be an undesirable feature, as consequences of electrolyte decomposition (such as cell dry-out) can proceed without obvious warning.

The reservoir effect is still largely at play in the phenomena described in the present work. Altering the DOD, for example, can carry the cell to a state in which two things can occur: i) the slope of the PE voltage profile changes at the EOD, causing $Li^+$ loss to raise the terminal PE potential in successive cycles by a different extent, which will also trigger an equal increase in NE potential, draining more or less of the NE reservoir than before; and/or ii) the slope of the NE profile changes at the EOD, altering the amount of "extra capacity" that is drained from the reservoir as a response to cell aging. Similar arguments could be made about changes in the depth of charge, which would affect the amount of $Li^+$ that can be extracted from the PE in future cycles as slippage causes both the PE and NE to reach the EOC at different terminal charge potentials over the cell's life. Simply put, variations in the shape of the voltage profiles will affect how well the measurements of coulombic efficiency and capacity will correspond to the actual rates of aging exhibited by the cell.

Naturally, the effects discussed above are intrinsic to Li-ion batteries, and can affect both how the extent of aging is perceived for a given cell and how one can interpret aging behavior across different cell chemistries and/or testing conditions. How can we neutralize this reservoir effect and enable more accurate analysis to be drawn from testing data?

On a more quantitative level, accurate determination of slippage is likely the most robust approach to gauge aging while evading the measurability issues caused by the reservoir effect.[5] Slippage is a direct consequence of the fact that side reactions will proceed independently at the PE (oxidation) and the NE (reduction), altering the individual *electrode SOCs* that correspond to a given cell *SOC*. Hence, slippage bears the signature of the net capacity that is exchanged in these parasitic processes. Consider again Figure 7a. A same extent of side reactions causes a same level of offset between the voltage profiles of the PE and the NE. That is, the extent of slippage depends only on the cumulative net parasitic capacity ($q_{red} - q_{ox}$), regardless of the details of the testing procedure (such as the DOD). However, out of this singular set of simulated aging, the varying perceived levels of capacity fade of Figure 7b are obtained once each voltage cutoff condition is imposed. This is a direct consequence of the reservoir effect. Depending on the shape of the voltage profiles of the PE and the NE around the EOD, the effective discharge profile of the aged cell could encompass a wider capacity range between the upper and lower cutoff voltages.[5] While directly *measuring* capacity would conceal the true extent of aging (Figure 7b), this information would be available from directly inferring the extent of slippage. Analytical techniques that can quantify slippage, such as differential voltage analysis,[10, 12] are extremely valuable to correctly gauge the consequences of cell aging.

On a more *qualitative* level, there are alternatives to improve the reliability of the analysis of aging trends. Once more, take the effect of DOD (Figure 6 and Figure 7) as example. Although the direct comparison of CE and CR measurements from cells tested under different voltage windows can be misleading, this can be solved by including occasional cycles that are performed within identical cutoff voltages. This option is already implemented in much of battery research, in which periodic reference performance tests (RPTs) including a common UCV and LCV are performed across all test conditions.[11, 13, 14] In this case, comparing the RPTs rather than the aging cycles can provide information about the relative effects of each testing protocol. The reservoir effect could remain an issue when RPTs cannot be implemented, such as in Li-ion batteries powering devices exposed to irregular charge and discharge regimes. In that case, for many battery formulations of interest, it could be difficult to determine the true extent of aging even if the total charge balance of the cell (i.e., charge vs. discharge exchanged capacity) can be perfectly tracked.

For cells employing different active materials in the PE and/or NE, it may be more difficult to perform a reliable comparative analysis, even a qualitative one. Ideally, there would always be a set of conditions at which all cells would provide equal measurability of aging; i.e., they would have similar $\lambda$ and $(1 + \omega - \lambda)$ values. In practice, this is often untrue and the equivalent of a RPT for different cell formulations may not exist. See for example Figures 6a,b. Although the differences in the measurability of aging among LFP and NMC cells vs. Gr reaches a minimum at full discharge, a finite gap still remains, which Figure 5 shows can produce sizable effects on perceived performance. In this case, using alternative analytical tools to help infer cell health, and/or analyzing slippage could be essential to identify whether the materials actually affect the rate of aging. This gap is even larger when attempting to compare NMC cells with Si-based materials versus ones with graphite (Figures 5 and 6).

Regardless of the systems under investigation, taking the reservoir effect into account is of the utmost importance when translating experimental data into conclusions about the true extent of aging endured by the cell.

*Conclusions*

Capacity and coulombic efficiency are generally considered reliable indicators of the instantaneous rate of aging experienced by a Li-ion battery. The present work discussed many instances in which this is *not* the case.

If we disregard impedance rise and losses of accessible active material capacity, cell aging is a consequence of the balance between reductive and oxidative side reactions. While reduction (SEI growth) will decrease the initial electron inventory of the cell, oxidation at the positive electrode will increase it. Since both parasitic processes occur at the expense of electrolyte decomposition and one of them does not deplete the electron inventory, capacity measurements are intrinsically flawed, to some extent, as a direct probe of the state of health of the cell. But even if this effect is small, how well can capacity and coulombic efficiency measurements convey the extent of parasitic processes? The answer depends on several factors.

Generally, the amount of capacity that is exchanged within a half-cycle is determined by the point where the difference between the instantaneous potentials of the positive and negative electrodes meet a specified cutoff voltage. Portions of the voltage profile of the PE and NE that are beyond this cutoff will not engage in capacity exchange at that half-cycle. Consequently, capacity measurements will only track a portion of the entire electron inventory that exists in the cell – there may be more $Li^+$ available in the PE at the end of charge, or the NE may still hold some $Li^+$ when the cell reaches the end of discharge. In other words, either electrode may contain a r*eservoir* of $Li^+$ that is unknown to capacity measurements. Slippage due to side reactions will cause a same cutoff voltage to occur at a different set of PE and NE potentials, altering the portions of the voltage profile of each electrode that are active within a given half-cycle. This change can cause part of the $Li^+$ reservoir to be added to the capacity that is exchanged by the cell as a consequence of aging. Thus, the true extent by which SEI growth will decrease the measurable cell capacity depends on how much extra $Li^+$ from the reservoir is introduced into the cell at each cycle.

The present work provided expressions that implicitly quantify this *reservoir effect*, by correlating the measurable CE and capacity retention values with the time-averaged rates of reduction and oxidation side reactions. Whether SEI growth can be captured by these measurements depends on the relationship between the slopes exhibited by the voltage profiles of the PE and the NE at the end of each half-cycle. Depending on the shape of the profiles, CE and CR measurements may actually provide little insight into the extent of aging of the cell, or may not inform at all about the rate of SEI growth. The expressions were then used to analyze the effect of the voltage profile of the PE active material and the depth of discharge on measurable aging outcomes. We demonstrate, for example, that a same rate of SEI growth will cause faster capacity fade in a cell using LFP vs. one using NMCs. Likewise, we show that a same extent of SEI growth in a NMC vs. graphite cell could generate more or less measurable capacity fade depending on the depths of charge and discharge used in the tests.

This work showed that distinguishing battery performance across testing conditions and electrode formulations can be more difficult than anticipated. Additionally, it proposed that certain pairings of positive and negative electrodes may provide higher levels of information about aging through direct measurements than others. Overall, all the examples we discuss here suggest that the true state of health of the cell may not always be correctly assessed through capacity measurements.

*Acknowledgements*

This research was supported by the U.S. Department of Energy's Vehicle Technologies Office under the Silicon Consortium Project, directed by Brian Cunningham, and managed by Anthony Burrell. The author is grateful to the colleague Adam Tornheim for his helpful comments during the preparation of this manuscript. The electrodes used in this article were fabricated at Argonne's Cell Analysis, Modeling and Prototyping (CAMP) Facility, which is fully supported by the DOE Vehicle Technologies Office (VTO). The submitted manuscript has been created by UChicago Argonne, LLC, Operator of Argonne National Laboratory ("Argonne"). Argonne, a U.S. Department of Energy Office of Science laboratory, is operated under Contract No. DE-AC02-06CH11357. The U.S. Government retains for itself, and others acting on its behalf, a paid-up nonexclusive, irrevocable worldwide license in said article to reproduce, prepare derivative works, distribute copies to the public, and perform publicly and display publicly, by or on behalf of the Government.

the preparation of this manuscript. The electrodes used in this article were fabricated at Argonne's Cell Analysis, Modeling and Prototyping (CAMP) Facility, which is fully supported by the DOE Vehicle Technologies Office (VTO). The submitted manuscript has been created by UChicago Argonne, LLC, Operator of Argonne National Laboratory ("Argonne"). Argonne, a U.S. Department of Energy Office of Science laboratory, is operated under Contract No. DE-AC02-06CH11357. The U.S. Government retains for itself, and others acting on its behalf, a paid-up nonexclusive, irrevocable worldwide license in said article to reproduce, prepare derivative works, distribute copies to the public, and perform publicly and display publicly, by or on behalf of the Government.

Table 1. Correlating the shapes of voltage profiles with measurement outcomes. For cells with PE-limited charge ($\omega \to 0$), the relative slopes of electrodes at the end of discharge will determine the type of information conveyed by coulombic efficiency and how much measurable capacity fade is produced by aging.

| Parameters | State | CE Information | CR Measurability |
|---|---|---|---|
| $\lambda \to 1$ and $\omega \to 0$ | NE-limited discharge (Figure 1a) | reduction side reactions | high |
| $\lambda \to 0$ and $\omega \to 0$ | PE-limited discharge (Figure 1b) | oxidation side reactions | how |
| $\omega \to 0$ and intermediate values of $\lambda$ | hybrid | reduction + oxidation side reactions | intermediate |

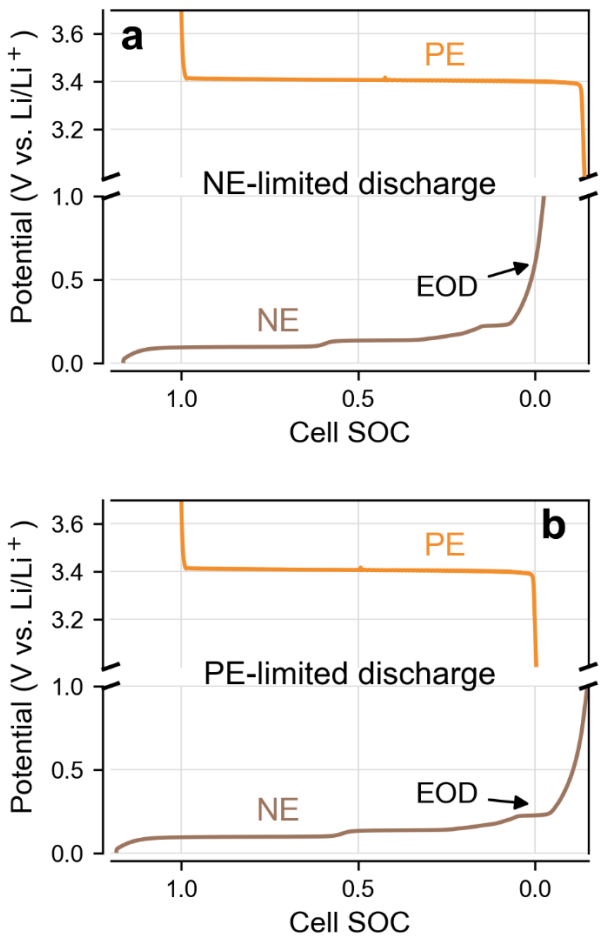

Figure 1. The role of electrodes in dictating the end of cell discharge. In theory, complete cell discharge can occur when: a) the negative electrode (NE) is virtually depleted of Li$^+$; or b) the positive electrode (PE) runs out of available sites to accommodate the incoming Li$^+$. In the latter case, SEI growth may not cause measurable capacity fade. Only portions of the voltage profiles of the LFP PE and the graphite NE that are in between the state of charge of 0 and 1 are actively utilized in the represented half-cycles.

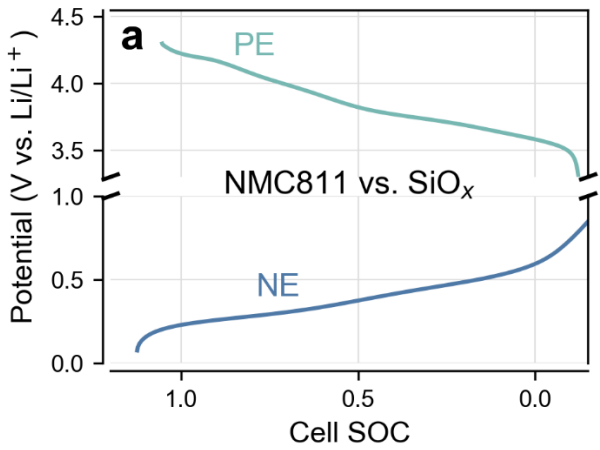

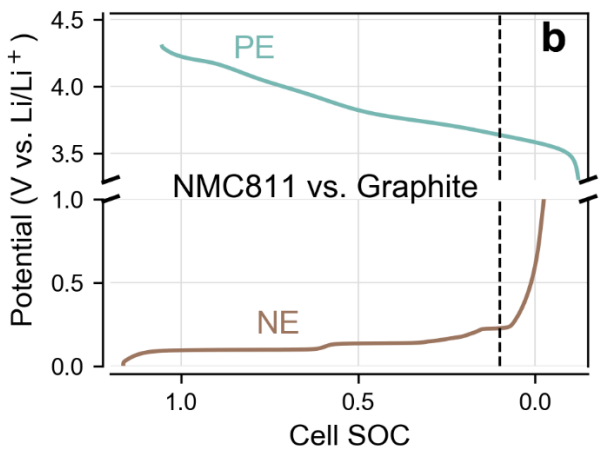

Figure 2. Indeterminacy of discharge limitation. a) Simulated voltage profiles during discharge of a NMC811 vs. $SiO_x$ cell. Profiles from both electrodes present similar slopes when the cell is fully discharged (to SOC = 0), causing the end of discharge to be limited by neither electrode. b) Simulated voltage profiles during discharge of a NMC811 vs. graphite cell. Although the cell is strongly NE-limited when fully discharged, interrupting discharge at SOC = 0.1 (dashed line) would cause the cell to be PE-limited. In both panels, full discharge would occur at an assumed cell voltage of 3 V.

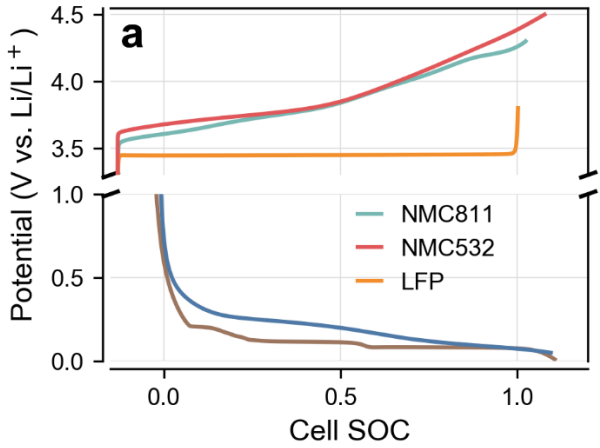

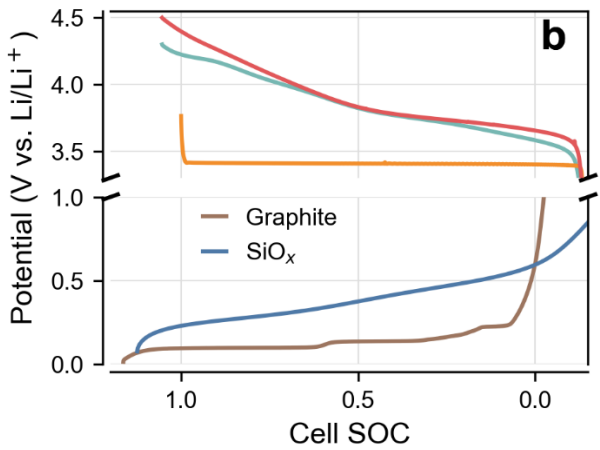

Figure 3. Voltage profiles of the positive and negative electrodes for hypothetical full-cells during: a) charge; and b) discharge. Six possible cells are represented, through the combination of the three types of PE materials with the two different NEs. Only portions of the voltage profiles that are in between the state of charge of 0 and 1 are actively utilized in the represented half-cycles. The legends apply to both panels.

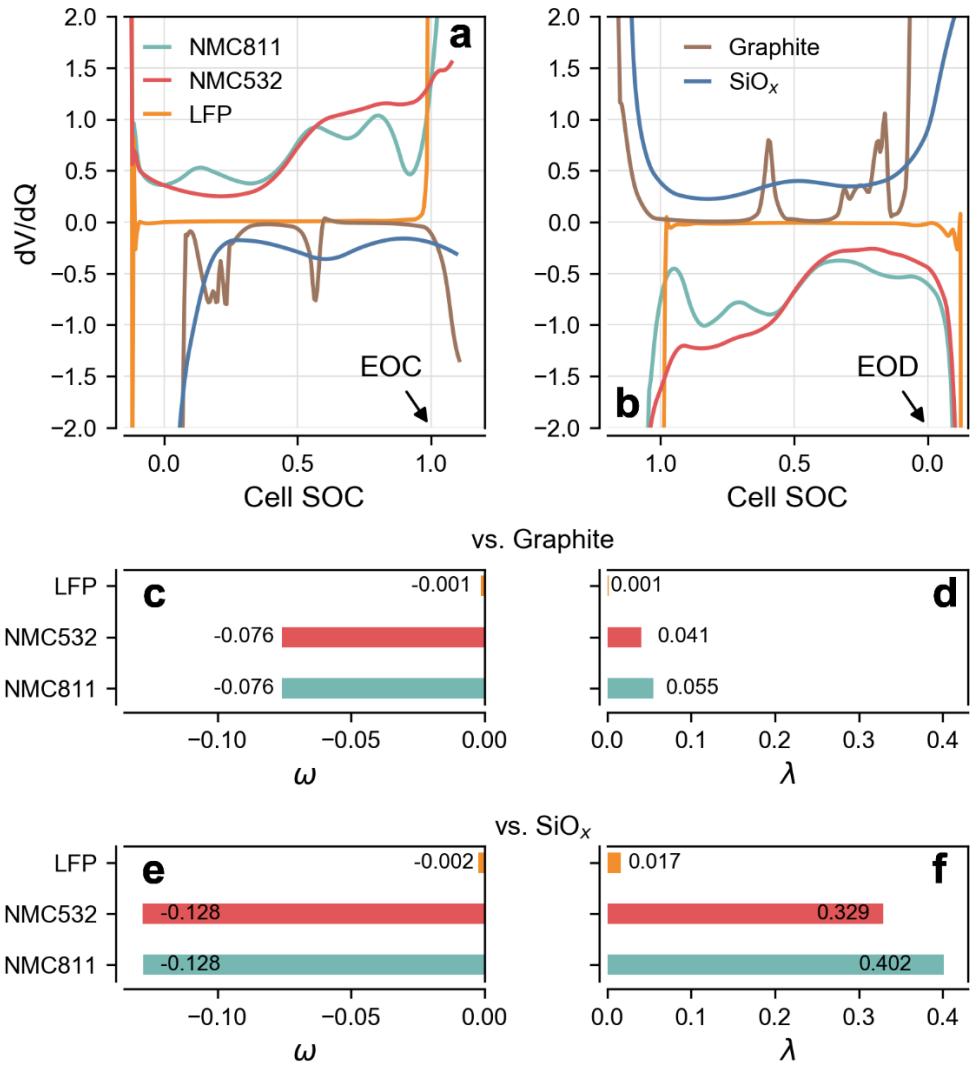

Figure 4. Quantifying the differences in the shape of voltage profiles. a) dV/dQ of the profiles from Figure 3a. b) dV/dQ of the profiles from Figure 3b. c) $\omega$ and d) $\lambda$ calculated for graphite cells vs. the indicated PE materials. e) $\omega$ and f) $\lambda$ calculated for SiO$_x$ cells vs. the various PEs. Refer to panels a and b for the color code.

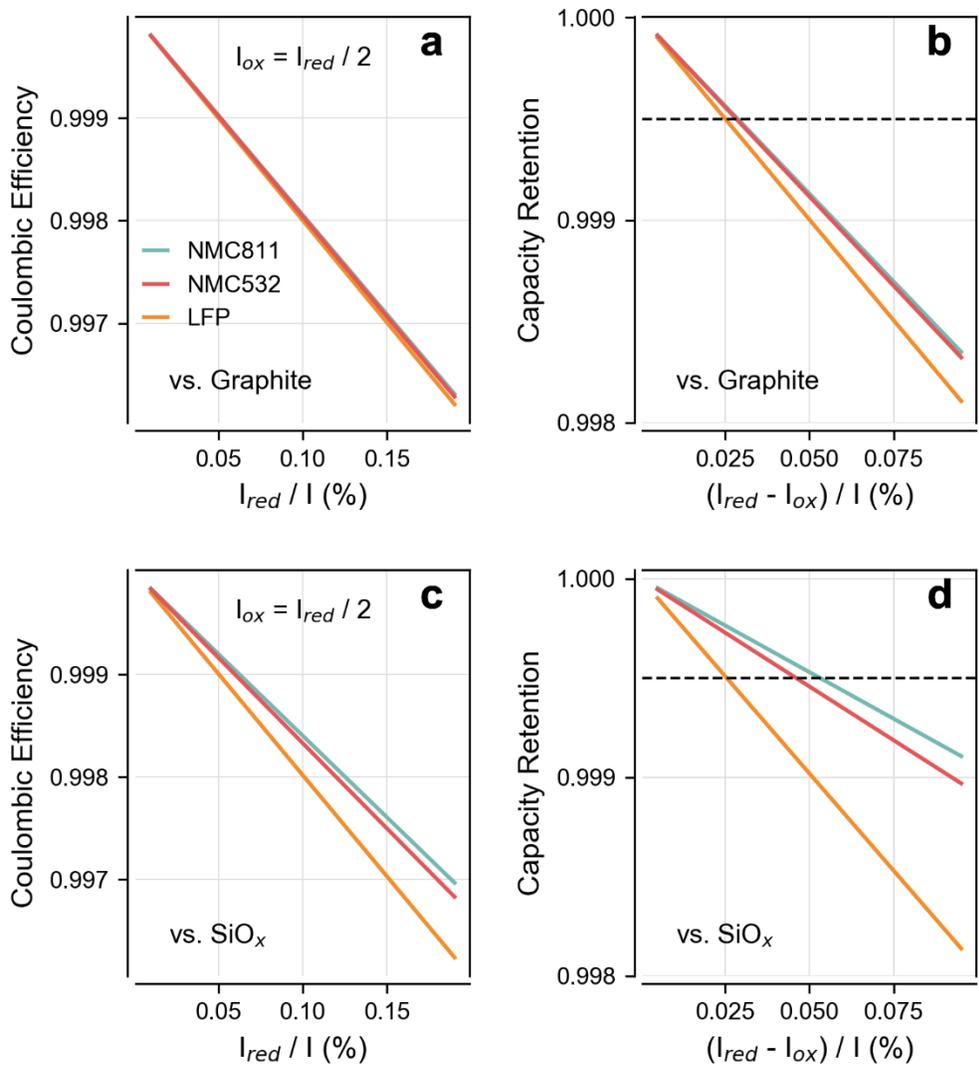

Figure 5. How the identity of the PE affects measurable quantities. a) coulombic efficiency and b) capacity retention for graphite cells versus the indicated PEs. c) coulombic efficiency and d) capacity retention for $SiO_x$ cells vs. the various PEs. A same rate of aging will result in lower CE and CR in LFP than in NMC cells. The black dashed line in panel b and d indicate a CR of 0.9995. Values were computed using $\lambda$ and $\omega$ shown in Figure 4. CE values were calculated

assuming that the rate of oxidation was half of the rate of reduction side reactions. The legend in panel a applies to all panels.

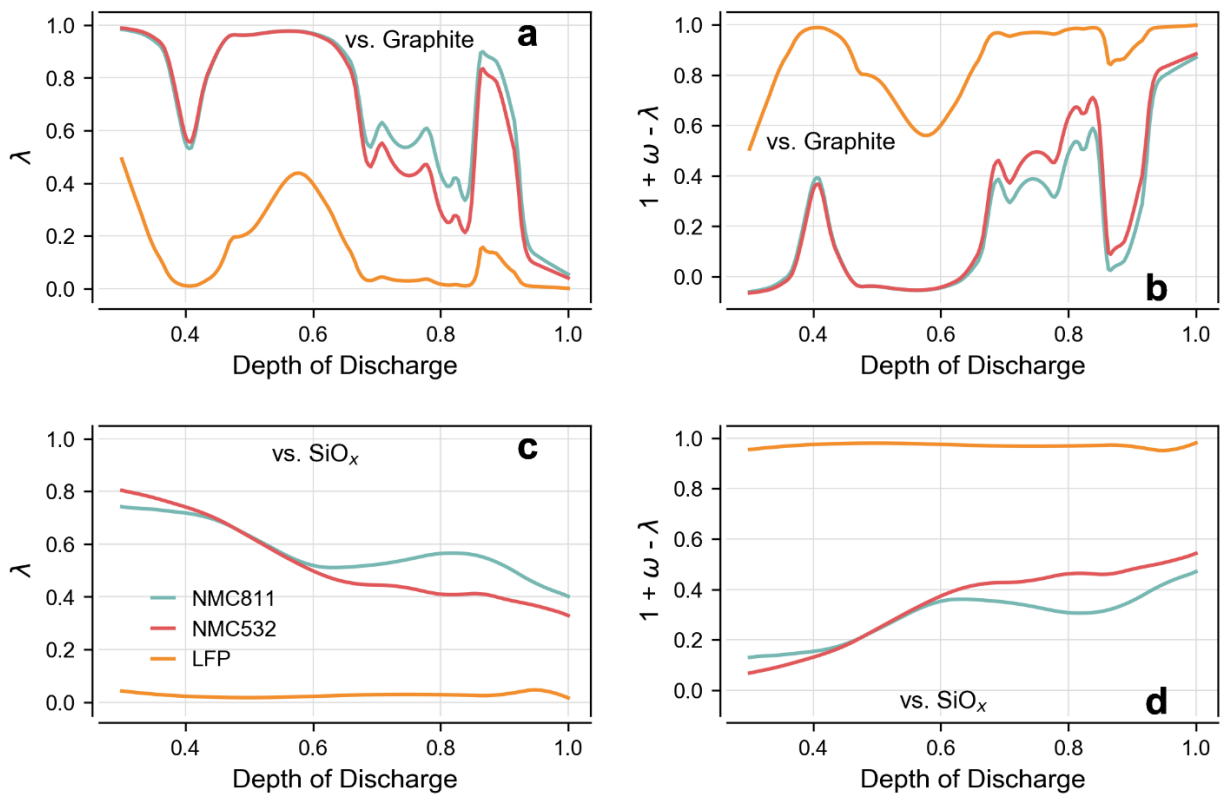

Figure 6. Parametrizing the effects of depth of discharge. a) $\lambda$ and b) $[1 + \omega - \lambda]$ for graphite cells vs. the indicated PEs. c) $\lambda$ and d) $[1 + \omega - \lambda]$ for $SiO_x$ cells vs. the various PEs. Cell discharge can switch from strongly NE- to strongly PE-limited depending on the voltage cutoff. Values were computed using the dV/dQ curves in Figure 4, assuming that all cells were previously fully charged. The legend in panel c applies to all panels.

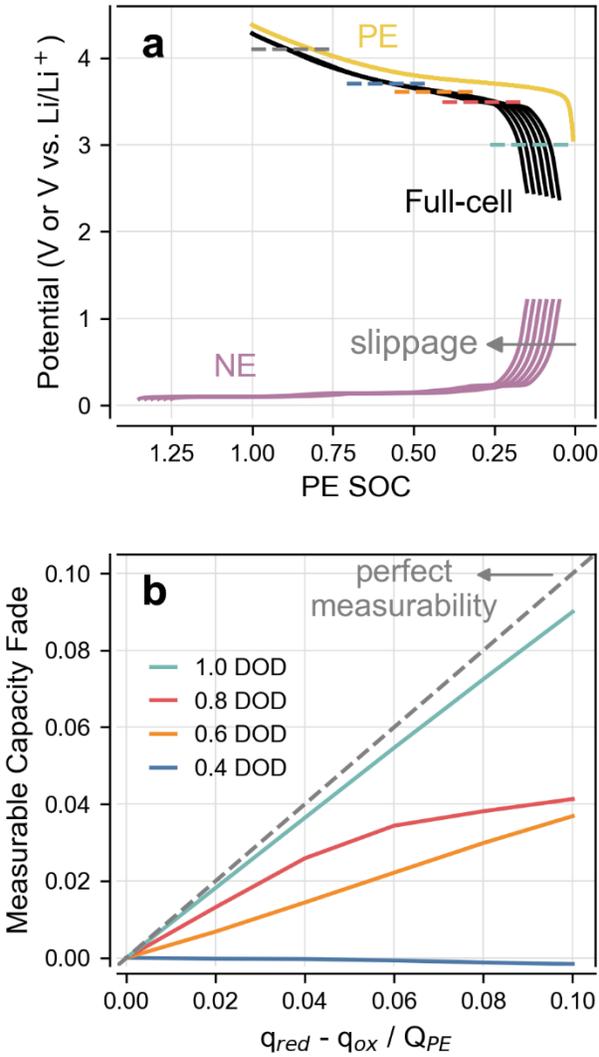

Figure 7. Simulating the effect of depth of discharge using the `Alawa toolbox. a) Voltage profiles of electrodes and full-cell during discharge of a NMC vs. graphite cell. Slippage is represented by the left-shift of the NE profiles with respect to the stationary voltage profile of the PE. Curves indicate slippages in steps of 2% of the PE SOC, to a maximum of 10%. The upper cutoff voltage is indicated by the gray dashed line, while the EOD voltages corresponding to the various DODs are marked by the colored ones. b) Cell capacity obtained after constraining the calculated full-cell voltage profiles within the UCV and each DOD. The x-axis shows the extent of slippage, indicated by the net charge lost by the cell normalized by the nominal capacity of the PE. All curves deviate from the diagonal line that indicates the case in which net electron consumption by parasitic processes is perfectly measurable.

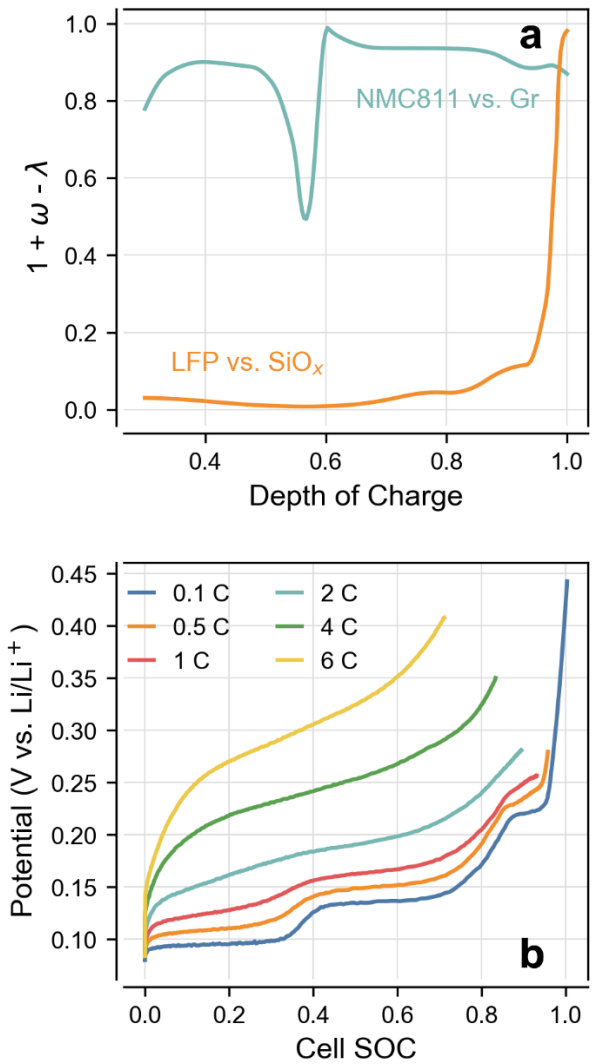

Figure 8. Additional examples affecting the measurability of aging. a) $[1 + \omega - \lambda]$ calculated for select cells as a function of the depth of charge. The ability of capacity measurements to track the net electron consumption by side reactions depends on the choice of upper cutoff voltage. It is assumed that all cells are fully discharged. b) NE profiles measured with a reference electrode during the discharge of a NMC532 vs. graphite cell at various rates. The shape of the voltage profile at all SOCs varies widely with the discharge current, affecting how CE and CR can convey aging information. Cell was fully charged prior to the half-cycles shown in the figure.